\pdfoutput=1
\documentclass[sigconf]{acmart}
 
\makeatletter                  
\def\mdseries@tt{m}             
\makeatother                  
\usepackage[plain]{fancyref} 
\usepackage[draft=true]{minted} 
\usepackage{color}
\usepackage{hyperref}           
\hypersetup{
    colorlinks=true,
    linkcolor=blue,
    filecolor=red,      
    urlcolor=magenta,
    breaklinks=true,            
}
\usepackage{breakurl} 

\AtBeginDocument{%
  \providecommand\BibTeX{{%
    \normalfont B\kern-0.5em{\scshape i\kern-0.25em b}\kern-0.8em\TeX}}}

\acmConference[GLSVLSI '23] {Great Lakes Symposium on VLSI 2023}{June 5-7, 2023}{Knoxville, TN, USA}

\renewcommand\footnotetextcopyrightpermission[1]{} 

\makeatletter

\usepackage[english]{babel}
\usepackage[utf8]{inputenc}
\usepackage{enumerate}
\usepackage{graphicx}
\usepackage{inputenc}
\usepackage[linesnumbered,ruled,vlined]{algorithm2e}
\usepackage{amsmath}
\usepackage{algorithmic}
\usepackage{upquote}
\usepackage{xcolor}
\usepackage{comment}
\usepackage{mathtools}
\usepackage{multirow}
\usepackage{lscape}
\usepackage{adjustbox}
\usepackage{fancyhdr}

\usepackage[utf8]{inputenc}
\usepackage{pgfplots}
\DeclareUnicodeCharacter{2212}{−}
\usepgfplotslibrary{groupplots,dateplot}
\usetikzlibrary{patterns,shapes.arrows}
\pgfplotsset{compat=newest}

\usepackage{subcaption}

\usepackage{tabularx}
\usepackage{dblfloatfix}

\usepackage{siunitx}

\newcommand*\CircledText[1]{\tikz[baseline=(char.base)]{
            \node[shape=circle,draw,inner sep=0.5pt] (char) {#1};}}
\begin{document}

\sloppy

\newcommand{\papername}{Locate}
\title{\papername{}: \texorpdfstring{\underline{Lo}w-Power Viterbi De\underline{c}oder Exploration using \underline{A}pproxima\underline{te} Adders}{Lg}}

\author{Rajat Bhattacharjya}
\affiliation{%
  \institution{TCS Research}
  \state{Mumbai, Maharashtra}
  \country{India}
}
\email{bhattacharjya.rajat@tcs.com}

\author{Biswadip Maity}
\affiliation{%
  \institution{University of California Irvine}
   \state{Irvine, CA}
   \state{USA}
}
\email{maityb@uci.edu}

\author{Nikil Dutt}
\affiliation{%
  \institution{University of California Irvine}
  \state{Irvine, CA}
  \country{USA}
}
\email{dutt@uci.edu}
\renewcommand{\shortauthors}{Bhattacharjya, et al.}


\begin{abstract}
Viterbi decoders are widely used in communication systems, natural language processing (NLP), and other domains. While Viterbi decoders are compute-intensive and power-hungry, we can exploit approximations for early design space exploration (DSE) of trade-offs between accuracy, power, and area. We present Locate, a DSE framework that uses approximate adders in the critically compute and power-intensive 
Add-Compare-Select Unit (ACSU) of the Viterbi decoder. We demonstrate the utility of Locate for early DSE of accuracy-power-area trade-offs for two applications: communication systems and NLP, showing a range of pareto-optimal design configurations. For instance, in the communication system, using an approximate adder, we observe savings of 21.5\% area and 31.02\% power with only 0.142\% loss in accuracy averaged across three modulation schemes. Similarly, for a Parts-of-Speech Tagger in an NLP setting, out of 15 approximate adders, 7 report 100\% accuracy while saving 22.75\% area and 28.79\% power on average when compared to using a Carry-Lookahead Adder in the ACSU. These results show that Locate can be used synergistically with other optimization techniques to improve the end-to-end efficiency of Viterbi decoders for various application domains.
\end{abstract}




\settopmatter{printfolios=true}
\maketitle
{
\renewcommand{\thefootnote}{\fnsymbol{footnote}}
\footnotetext{Paper accepted at ACM GLSVLSI 2023.
}
\footnotetext{ Authors' version of the work posted for personal use and not for redestribution. The definitive version will be available in the Proceedings of the ACM GLSVLSI 2023 at https://doi.org/10.1145/3583781.3590314. }
}
\section{Introduction}
\label{sec:intro}
The Viterbi decoder has been widely applied in diverse fields - including Natural Language Processing, Speech Recognition, and modern communication systems such as LTE PDCCH, WLAN, CDMA, and satellite communication.
Over time, there have been consistent efforts to study and improve upon Viterbi decoders~\cite{resilient, bougard, vit1, vit2, nitin} given their wide range of applications across various domains.

Viterbi decoders are known to be computationally expensive and challenging to implement in low-power hardware. 
We look at two examples:
(1) \textit{Digital Communication Systems}: The increase in power of Viterbi decoders is particularly pronounced in communication systems where channel conditions are impaired~\cite{resilient}. 
Studies have shown that even under ideal conditions, such as those mandated by IEEE 802.11 a/g WLAN standards, a Viterbi decoder, with 128 states and operating at 54 Mbps, consumes 35\% of the transceiver's total power~\cite{bougard} (including the digital and analog front ends). In terms of operations per second, this accounts for 76\% of all digital processing complexity~\cite{recon}. 
(2) \textit{Natural Language Processing (NLP)}: Viterbi decoding is challenging in NLP due to the large number of sequences to consider, the complexity of language, and ambiguity in determining the most likely sequence. This results in high computational complexity and power consumption.
As a result, it has become ever-important to propose methods to lower power consumption in such decoders in order to make them more practical for NLP tasks.

\begin{figure}
\includegraphics[width=0.45\textwidth]{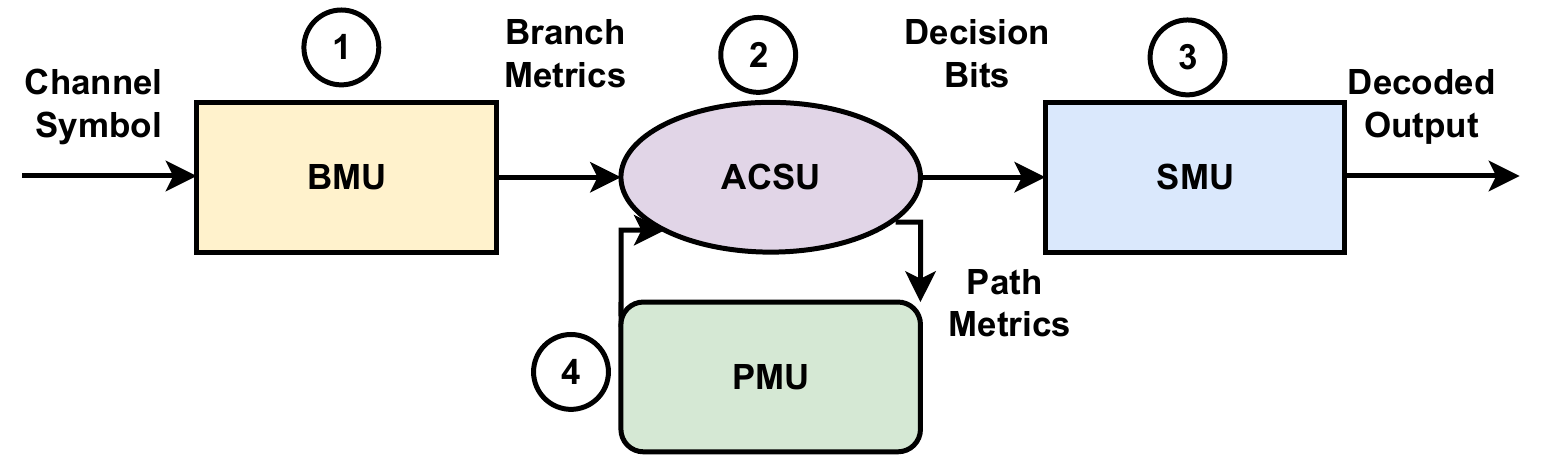}
\caption{Functional diagram of a Viterbi decoder.}
\label{fig:vit}
\vspace{-2ex}
\end{figure}

Previous works have proposed various methods to reduce the power consumption of Viterbi decoders in hardware. 
Abdallah et al.~\cite{resilient} proposed two low-power decoders that exploit error resilience. 
Hegde et al.~\cite{err} used Voltage Overscaling (VOS) and algorithmic noise tolerance (ANT) to reduce power and correct timing errors.
Other works have proposed reducing state sequence decoder~\cite{seq}, survivor memory size~\cite{surv}, and limited search trellis~\cite{limit}, all resulting in a trade-off between performance and accuracy. 
In this work, we focus on reducing power consumption by  exploiting approximation in adders, 
a critical component in the most compute-intensive parts of a Viterbi decoder.

Specifically our \textit{\papername{}}  design space exploration (DSE) approach 
utilizes
approximate adders in the Add-Compare-Select-Unit (ACSU) of the Viterbi decoder (shown as unit \CircledText{2} in Figure \ref{fig:vit}, explained further in Section \ref{sec:background}).
This ACSU is known to be a compute-intensive kernel that results in high hardware complexity and power consumption, particularly with increased constraint lengths as it involves a large number of trellis states and transitions~\cite{acs}. 
By introducing approximations in this unit, 
\textit{\papername{}} enables the exploration of pareto-optimal accuracy-power-area design points,
allowing system designers to identify interesting design configurations that reduce
the power consumption of a Viterbi decoder while maintaining a reasonable level of accuracy.
%
%
We obtain optimal design points through experimental evaluation and validation of the proposed approach.
For a digital communication system, the generated low-power decoder with the adder add12u\_187~\cite{evoapprox} saves 21.5\% area, and 31.02\% power (averaged across three modulation schemes, BASK, BPSK, and QPSK) compared to a decoder with Carry-Lookahead Adder (CLA) (in the ACSU) with a 0.142\% loss in accuracy.  For a Parts-of-Speech Tagger in an NLP setting, we observe that out of 15 approximate adders, 7 report 100\% accuracy while saving 22.75\% area and 28.79\% power on average compared to a decoder with CLA (in the ACSU).



\section{Background}
\label{sec:background}
The Viterbi decoder, as shown in Figure~\ref{fig:vit}, is a widely used component for decoding convolutional or trellis codes. 
The underlying algorithm, known as the Viterbi algorithm, employs dynamic programming techniques and is frequently used with Markov sources and hidden Markov models (HMM)~\cite{viterbi}. 
The Viterbi algorithm estimates the probability of the most likely sequence of hidden states, referred to as the Viterbi path. 
The Viterbi decoding method is widely used in various applications, including digital communication, natural language processing (NLP), speech recognition, and DNA analysis.

The four key units of the Viterbi decoder (Figure~\ref{fig:vit}) are as follows:
The Branch Metric Unit (BMU) \textsuperscript{\CircledText{1}} is a device that calculates the distance between each symbol in a code alphabet and the received symbol, utilizing a norm to standardize the distance. These standardized distances are referred to as Branch Metrics (BMs). 
The BMs are fed into the Add-Compare-Select Unit (ACSU) \textsuperscript{\CircledText{2}}, which recursively calculates the Path Metrics (PMs) and produces decision bits as output for each possible state transition.
The decision bits are subsequently stored and retrieved from the Survivor Memory Unit (SMU) \textsuperscript{\CircledText{3}}, in order to obtain the decoded source bits along the ultimate survivor path. 
PMs for the present iteration is stored in the Path Metric Unit (PMU)\textsuperscript{\CircledText{4}}.

Given the error-resilient nature of the Viterbi algorithm and its wide range of applications, 
it is amenable to
exploration of approximate computing techniques to further reduce power consumption and increase efficiency. 
Approximate computing has evolved as a popular alternative to conventional computing approaches given the error-resilient nature of algorithms and applications~\cite{app}.
Approximations can be applied at different abstraction levels starting from full-system level~\cite{raha} to software level~\cite{paraprox} and storage~\cite{mem, 10.1145/3466875}. 
Arithmetic units such as adders and multipliers form the basic building blocks in all such approximate systems~\cite{cesa, morteza, rearm, acla}. 



Having identified the ACSU as the power bottleneck for Viterbi decoders,
our \papername{} approach complements  
prior optimization efforts
by deploying approximate adders at the hardware level within the power-hungry ACSU,
and exploring the accuracy-power-area design space
with applications in digital communication and NLP.
\section{Methodology}
\label{sec:method}
 \begin{figure}
\includegraphics[width=0.48\textwidth]{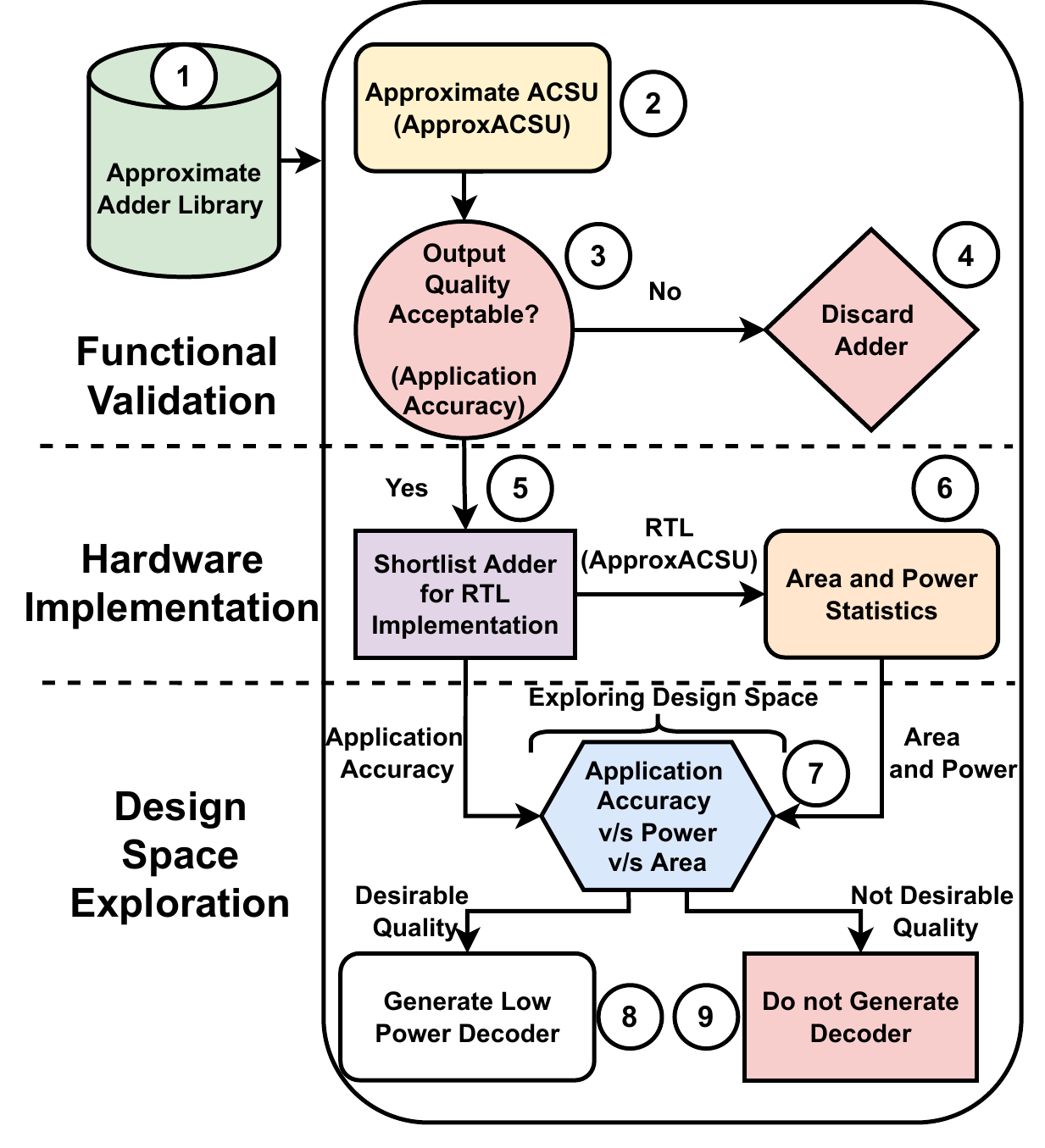}
\caption{Methodology of Locate}
\label{fig:method}
\vspace{-3ex}
\end{figure}

Figure~\ref{fig:method} outlines our \textit{\papername{}} methodology for Low Power Decoder Exploration of Viterbi decoders. 
It comprises three main steps: Functional Validation at the Software Level, Hardware Implementation, and Design Space Exploration to generate a decoder. 
The steps are described below:

\begin{figure*}
    \begin{center}
    \includegraphics[width=1\textwidth]{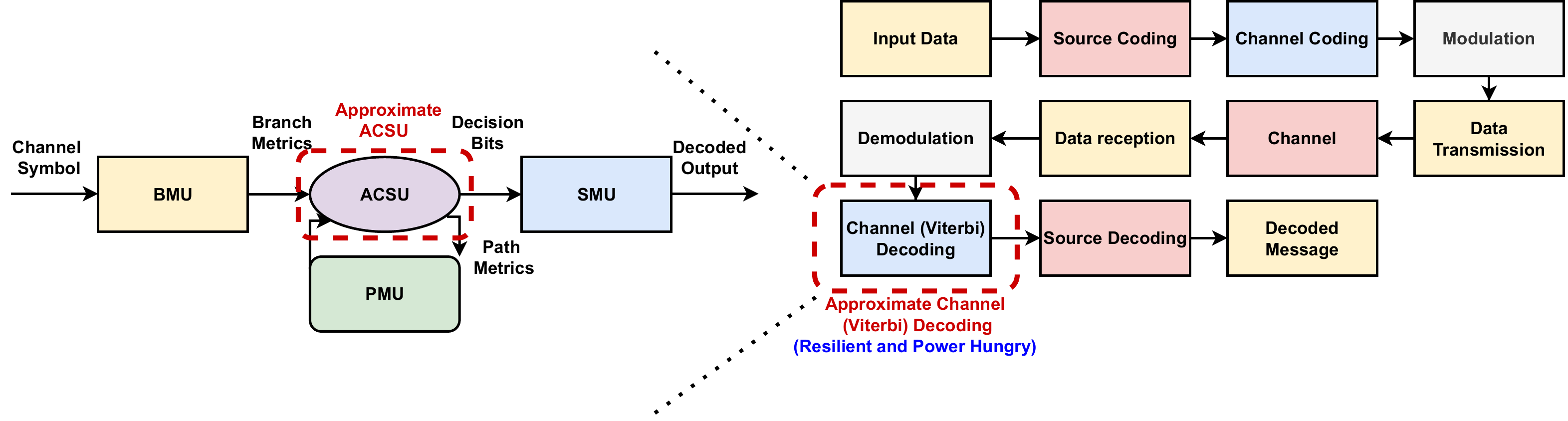}
    \caption{End-to-end Digital Communication System: Approximated units are highlighted in red.}
    \label{fig:overview}
    \end{center}
\end{figure*}
\subsection{Functional Validation}

Functional Validation is conducted at the software level using MATLAB/Python to evaluate the impact of approximations on the quality of the output.
A 
key question 
is the identification of the appropriate block to approximate and the reasoning behind this choice, within the constraints of the application's quality window.

Previous research has shown that despite being power hungry, Viterbi decoders are error-resilient~\cite{err}, making them perfect candidates for approximation techniques. 
However, it is necessary first to identify which specific unit within the Viterbi decoder (depicted previously in Figure~\ref{fig:vit}) is most amenable to approximation. 
We select the Add-Compare-Select Unit (ACSU) within the Viterbi decoder for this purpose.
The ACSU calculates PMs in a recursive manner and thereby gives out decision bits as output for each possible state transition. 
Approximating operations within the ACSU is advantageous as it reduces a significant amount of the complex computation associated with recursion.

In this paper, we investigate two primary applications that deploy Viterbi decoders: a complete end-to-end digital communication system and a Parts-of-Speech Tagger in the field of NLP. 
To determine the effectiveness of this approach in the specific context of \papername{}, an accuracy analysis per application is carried out with different hardware implementations.
This analysis involves approximating the addition operations within the ACSU\textsuperscript{\CircledText{2}} using various approximate adders sourced from EvoApprox library\textsuperscript{\CircledText{1}}~\cite{evoapprox}. 
If the output quality of the application meets the desired accuracy requirements\textsuperscript{\CircledText{3}}, the adders from the library are considered for implementation in RTL; otherwise, they are eliminated from further consideration\textsuperscript{ \CircledText{4}}.
It is important to note that not all adders are suitable for all applications, as they have different impacts on the accuracy metrics.
Some of them perform better than others, and therefore, a comprehensive design space exploration is necessary in order to evaluate the various accuracy-hardware trade-offs associated with this methodology.

In order to evaluate the accuracy of the approximate adders, a variety of metrics are considered, such as Mean Absolute Error (MAE), Error Percentage (EP), Worst-Case Absolute Error (WCE), Mean Squared Error (MSE), Mean Relative Error (MRE) etc.~\cite{stat}. 
It is important to note that various factors, including the distribution of input data and the digital design, can affect these errors. 
Thus, it is crucial to investigate these factors in order to gain a deeper understanding of which circuits are suitable for which applications and why.

\subsection{Hardware Implementation}
\label{method:hw}
To fully comprehend the benefits of approximations in terms of power and area savings, it is essential to conduct hardware implementation \textsuperscript{\CircledText{5}} of the identified blocks. 
Toward this, we implement the RTL of the ACSU using Verilog HDL, incorporating various approximate adders into the ACSU. 
We use Synopsys Design Compiler (DC) to synthesize both the approximate and the accurate ACSU designs with the 45nm NanGate Open Cell Library. 
By doing this, we can obtain area and power statistics \textsuperscript{\CircledText{6}} associated with these designs.
The approximate adders used to approximate the addition operations in the ACSU were selected from the EvoApprox Library~\cite{evoapprox}.
Next, we explore the design space of Viterbi decoders by varying the approximate adders. 

\subsection{Design Space Exploration}
Design Space Exploration (DSE) is a methodology used in electronic design automation (EDA) to identify the best design point for a system, considering various design constraints and objectives. 
In \papername{}, we are conducting a DSE of Viterbi decoders by exploring the trade-offs between accuracy, power, and area.

Having previously determined both accuracy and hardware (area, power) statistics, it is crucial for us to identify the pareto-optimal design points considering the trade-offs for a design point. 
A pareto-optimal design point is one that satisfies all design constraints and objectives, such as meeting a certain accuracy threshold while minimizing power consumption and area utilization.
Towards this end, we perform the design space exploration of Viterbi decoders (accuracy v/s power v/s area) \textsuperscript{\CircledText{7}} using various approximate ACSUs to obtain the pareto-optimal design points \textsuperscript{\CircledText{8}}. 
If the evaluation of the accuracy, power, and area trade-off is not deemed satisfactory, the adder corresponding to the design point is not used to generate a decoder\textsuperscript{\CircledText{9}}. 
It is important to note that this selection \CircledText{9} is separate from the selection in \CircledText{4}.
\CircledText{4} allows an early selection just after functional validation at the software level, whereas for  \CircledText{9}, the hardware implementation and design space exploration must also be conducted before reaching a conclusion.
This two-step approach enables us to identify the optimal Low-Power decoder design by filtering out adders that do not meet the desired criteria.
\section{Evaluation}
\label{sec:eval}
In order to assess the efficacy of approximating the addition operations within the Viterbi Decoder's ACSU, we evaluate two state-of-the-art applications: an end-to-end digital communication system (in Section \ref{sec:eval:comm}) and a Parts-of-Speech tagger, a key task in NLP (in Section \ref{sec:eval:nlp}).

The methodology employed for both of these applications remains the same (explained in Section \ref{sec:method}).
First, we perform the functional validation at the software level, which helps us validate the approach of approximating the addition operations inside the ACSU of the Viterbi decoder and thereby consider it in an end-to-end system.
Second, we perform the hardware evaluation of the impacted ACSU blocks using various approximate adders.
Finally, we perform a Design Space Exploration to find out pareto-optimal designs.

\begin{figure*}[b!]
\includegraphics[width=\textwidth]{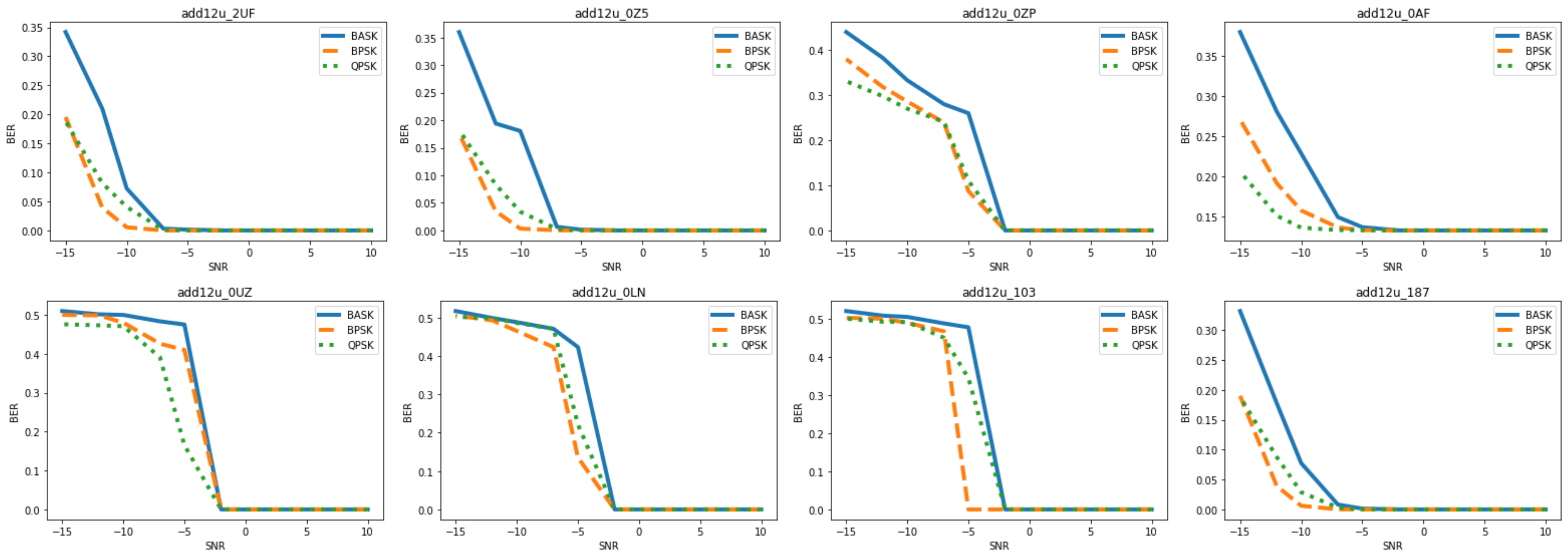}
\caption{BER vs SNR (text data transmission and reception) using various approximate adders in the Viterbi decoder for a complete digital communication system}
\label{fig:comm_acc}
\end{figure*}

\subsection{Digital Communication System}
\label{sec:eval:comm}
The digital communication system under consideration is shown in Figure~\ref{fig:overview}. 
The system is sourced from the repository provided in~\cite{mahmud}. 
The techniques used in the different modules of the communication system are outlined in Table~\ref{tab:comm_sys}, and the system properties for the entire model are highlighted in Table~\ref{tab:prop}.
Only the Channel Decoder block (the Viterbi decoder) is subjected to approximation, while the remaining blocks remain intact. 
An end-to-end performance analysis is conducted as a result of approximations in the channel decoder.

A total of 14 approximate adders are evaluated from the EvoApprox Library~\cite{evoapprox} to perform the approximation of the addition operations in the ACSU.
The adders are: add12u\_2UF, add12u\_39N, add12u\_0UZ, add12u\_0Z5, add12u\_0LN, add12u\_187, add12u\_0ZP, add12u\_103, add12u\_0AF, add12u\_28B, add12u\_4NT, add12u\_50U, add12u\_0C9, add12u\_0AZ.
We evaluate across 3 modulation schemes, namely, Binary Amplitude Shift Keying (BASK), Binary Phase Shift Keying (BPSK), and Quadrature Phase Shift Keying (QPSK).

\begin{table}
\caption{Communication system blocks and their properties}
\label{tab:comm_sys}

\begin{tabularx}{0.46\textwidth}{|l|X|}
 \hline
 \textbf{Block type} & \textbf{Technique Used} \\
  \hline
 Source Coding & Huffman Encoding \\
  \hline
 Channel Coding & Convolutional Encoding \\
  \hline
 Modulation & Amplitude Shift Keying (ASK) and Phase Shift Keying (PSK) \\
  \hline
 Channel & Additive White Gaussian Noise Channel \\
  \hline
 Demodulation & Amplitude Shift Keying (ASK) and Phase Shift Keying (PSK) demodulation \\
  \hline
 Channel Decoding &  Viterbi Decoding \\
  \hline
 Source Decoding & Huffman Decoding \\
  \hline
\end{tabularx}
\vspace{-3ex}
\end{table}

\subsubsection{Accuracy Analysis}
To evaluate the accuracy of the design points in a digital communication system, we transmit text data (653 words) and calculate any errors that appear upon receiving it. 
The errors are reported in terms of Bit Error Rate (BER), and the results of BER versus SNR (dB) of the communication system using 8 adders from the EvoApprox Library is shown in Figure~\ref{fig:comm_acc}, where SNR is X-axis and BER is Y-axis. The BER calculated per SNR is averaged across a dozen runs for each adder and modulation scheme.

We make 4 major observations. 
First, among the 14 adders, 6 were found to result in complete data corruption and were deemed unsuitable for use in the communication model (hence not shown in Figure~\ref{fig:comm_acc}). 
Second, add12u\_187 has the lowest BER averaged across all SNR levels. 
This is interesting because it has an MAE of 0.24\% and EP of 49.22\%~\cite{evoapprox} as compared to add12u\_2UF, which has an MAE and EP of 0\%. 
Third, compared to the accurate case, a marginal accuracy loss of 0.142\% in BER is reported (averaged across modulation schemes) when add12u\_187 is used. 
This supports the feasibility of using approximate adders for building low-power Viterbi decoders. 
Fourth, BPSK has lower BER than QPSK and BASK for most cases under negative SNR, but QPSK performs better in some cases. BASK has higher BER than PSK schemes in the negative SNR region, which is expected.

Finally, we conclude that though a particular circuit might have its own error metrics associated with itself, across an end-to-end system, using \papername{}, we observe behavior that is different than expected due to dynamic interactions between various modules. 

\begin{table}[t]
\caption{System Properties of the Communication Model}
\label{tab:prop}
\begin{tabular}{|c|c|}
 \hline
 \textbf{Property} & \textbf{Type} \\ \hline
 Modulation & BASK, BPSK, QPSK \\ \hline
 Samples per bit & 40 \\ \hline
 Bit rate & 1000 \\ \hline
 Carrier frequency & \SI{1000}{\hertz}\\ \hline
 Carrier Amplitude & 1 V\\ \hline
 SNR & -15 to 10 dB\\ \hline
 Generator Matrix for Channel Coder & {[}1 1 1; 1 0 1{]} \\ \hline
 Number of bit shift in register & 1 \\ \hline
\end{tabular}
\vspace{-3ex}
\end{table}

\setcounter{figure}{5}
\begin{figure*}[b!]
\begin{subfigure}[t]{0.3\textwidth}
    \centering
    \includegraphics[width=\textwidth]{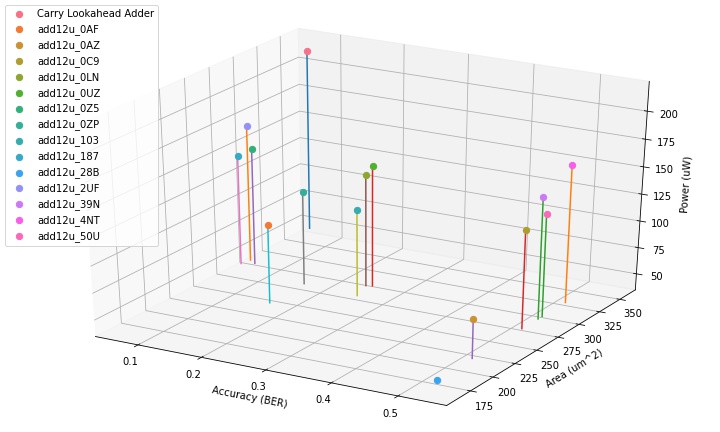}
    \caption{DSE for BASK}
    \label{fig:dse_bask}
\end{subfigure}%
\begin{subfigure}[t]{0.3\textwidth}
    \centering
    \includegraphics[width=\textwidth]{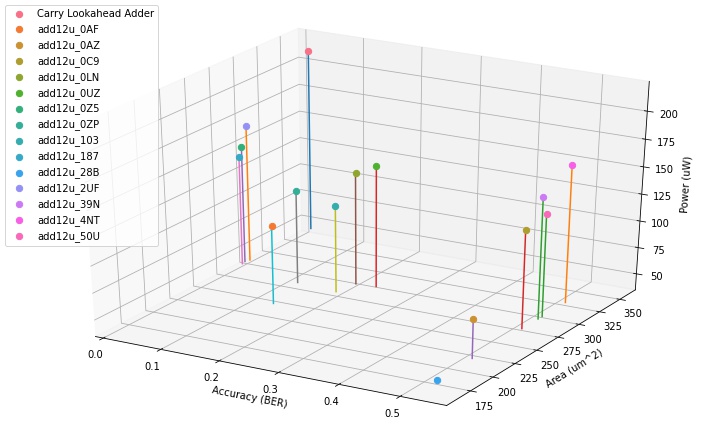}
    \caption{DSE for BPSK}
    \label{fig:dse_bpsk}
\end{subfigure}%
\begin{subfigure}[t]{0.3\textwidth}
    \centering
    \includegraphics[width=\textwidth]{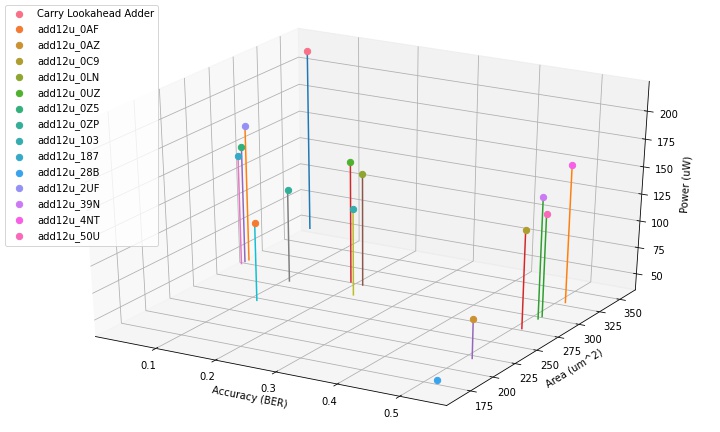}
    \caption{DSE for QPSK}
    \label{fig:dse_qpsk}
\end{subfigure}%
\caption{Design Space Exploration of Viterbi decoders for Digital Communication System using various adders and three modulation schemes: BASK, BPSK, QPSK}
 \label{fig:dse_comm} 
\end{figure*}

\subsubsection{Hardware Evaluation}
\label{sec:eval:comm:hw}

\setcounter{figure}{4}
\begin{figure}
    \centering
\begin{tikzpicture}

\definecolor{darkgray176}{RGB}{176,176,176}
\definecolor{gray}{RGB}{128,128,128}
\definecolor{lightgray204}{RGB}{204,204,204}

\begin{axis}[
legend cell align={left},
legend style={
  fill opacity=0.8,
  draw opacity=1,
  text opacity=1,
  at={(1.1,1.05)},
  anchor=east,
  draw=lightgray204
},
tick align=outside,
tick pos=left,
x grid style={darkgray176},
xmin=0, xmax=373.2351,
xtick style={color=black},
width=0.35\textwidth,
height=7cm,
y grid style={darkgray176},
ymin=-0.2, ymax=16,
ytick style={color=black},
ytick={0.4,1.4,2.4,3.4,4.4,5.4,6.4,7.4,8.4,9.4,10.4,11.4,12.4,13.4,14.4,15.4},
yticklabels={
  Ripple Carry Adder,
  CLA,
  add12u\_2UF,
  add12u\_39N,
  add12u\_0UZ,
  add12u\_0Z5,
  add12u\_0LN,
  add12u\_187,
  add12u\_0ZP,
  add12u\_103,
  add12u\_0AF,
  add12u\_28B,
  add12u\_4NT,
  add12u\_50U,
  add12u\_0C9,
  add12u\_0AZ
}
]
\draw[draw=none,fill=black] (axis cs:0,-0.2) rectangle (axis cs:291.802,0.2);
\addlegendimage{ybar,ybar legend,draw=none,fill=black}
\addlegendentry{Area ($\mu m^2$)}

\draw[draw=none,fill=black] (axis cs:0,0.8) rectangle (axis cs:355.462,1.2);
\draw[draw=none,fill=black] (axis cs:0,1.8) rectangle (axis cs:287.014,2.2);
\draw[draw=none,fill=black] (axis cs:0,2.8) rectangle (axis cs:277.438,3.2);
\draw[draw=none,fill=black] (axis cs:0,3.8) rectangle (axis cs:281.694,4.2);
\draw[draw=none,fill=black] (axis cs:0,4.8) rectangle (axis cs:283.29,5.2);
\draw[draw=none,fill=black] (axis cs:0,5.8) rectangle (axis cs:280.09,6.2);
\draw[draw=none,fill=black] (axis cs:0,6.8) rectangle (axis cs:279.03,7.2);
\draw[draw=none,fill=black] (axis cs:0,7.8) rectangle (axis cs:265.734,8.2);
\draw[draw=none,fill=black] (axis cs:0,8.8) rectangle (axis cs:262.542,9.2);
\draw[draw=none,fill=black] (axis cs:0,9.8) rectangle (axis cs:227.43,10.2);
\draw[draw=none,fill=black] (axis cs:0,10.8) rectangle (axis cs:163.05,11.2);
\draw[draw=none,fill=black] (axis cs:0,11.8) rectangle (axis cs:309.89,12.2);
\draw[draw=none,fill=black] (axis cs:0,12.8) rectangle (axis cs:282.226,13.2);
\draw[draw=none,fill=black] (axis cs:0,13.8) rectangle (axis cs:258.552,14.2);
\draw[draw=none,fill=black] (axis cs:0,14.8) rectangle (axis cs:202.16,15.2);
\draw[draw=none,fill=gray] (axis cs:0,0.2) rectangle (axis cs:176.435,0.6);
\addlegendimage{ybar,ybar legend,draw=none,fill=gray}
\addlegendentry{Power ($\mu W$)}

\draw[draw=none,fill=gray] (axis cs:0,1.2) rectangle (axis cs:214.044,1.6);
\draw[draw=none,fill=gray] (axis cs:0,2.2) rectangle (axis cs:172.4556,2.6);
\draw[draw=none,fill=gray] (axis cs:0,3.2) rectangle (axis cs:158.1465,3.6);
\draw[draw=none,fill=gray] (axis cs:0,4.2) rectangle (axis cs:158.44,4.6);
\draw[draw=none,fill=gray] (axis cs:0,5.2) rectangle (axis cs:154.6257,5.6);
\draw[draw=none,fill=gray] (axis cs:0,6.2) rectangle (axis cs:149.936,6.6);
\draw[draw=none,fill=gray] (axis cs:0,7.2) rectangle (axis cs:147.6282,7.6);
\draw[draw=none,fill=gray] (axis cs:0,8.2) rectangle (axis cs:132.304,8.6);
\draw[draw=none,fill=gray] (axis cs:0,9.2) rectangle (axis cs:126.5126,9.6);
\draw[draw=none,fill=gray] (axis cs:0,10.2) rectangle (axis cs:118.8094,10.6);
\draw[draw=none,fill=gray] (axis cs:0,11.2) rectangle (axis cs:46.76,11.6);
\draw[draw=none,fill=gray] (axis cs:0,12.2) rectangle (axis cs:172.6576,12.6);
\draw[draw=none,fill=gray] (axis cs:0,13.2) rectangle (axis cs:141.466,13.6);
\draw[draw=none,fill=gray] (axis cs:0,14.2) rectangle (axis cs:137.3639,14.6);
\draw[draw=none,fill=gray] (axis cs:0,15.2) rectangle (axis cs:82.731,15.6);
\end{axis}

\end{tikzpicture}
    \caption{Area and Power Statistics for Digital Communication System using various adders in the Viterbi Decoder}
    \label{fig:ap_comm} 
\end{figure}
\setcounter{figure}{6}

We synthesize the hardware using Synopsys DC as described in Section~\ref{method:hw} and obtain the comprehensive area and power statistics as shown in Figure~\ref{fig:ap_comm}. 
We observe that CLA consumes the most amount of area and power, which is trivial given the fact that the carry-lookahead circuit is complex in nature. The statistics also show that add12u\_28B consumes 
the least
area and power among all the adder candidates. 
However, it is important to note that add12u\_28B results in complete data corruption, and because of that reason, it is not feasible to use it in such an application. 
Next, the hardware statistics are used for the DSE.


\subsubsection{Design Space Exploration (DSE)}
\label{sec:comm_dse}
The relationships between accuracy, area, and power are complex:
some adders might have exceptional accuracy, but may suffer at the hardware level consuming more on-chip area and power; 
others might have excellent on-chip area and power statistics but extremely degraded application accuracy.
Thus it is important for us to comprehensively explore the design space involving a variety of factors.

For our experiments, we explore a 3D 
design space 
with the parameters: application accuracy (BER), on-chip area, and power consumed.
The BER observed is of the end-to-end digital communication model using three modulation schemes: BASK, BPSK, and QPSK. 
The area and power statistics are those of the ACSUs with various accurate and approximate adders. 
The results are shown in Figure~\ref{fig:dse_comm}.

$\bullet$ \textbf{BASK:} Figure~\ref{fig:dse_bask} shows the design space for BASK. We observe that though the BER for CLA is minimal, the area and power consumption are very high. 
If the required BER is <0.2, we see 6 candidates fulfilling that criterion. For an area budget <250 $\mu m^2$, we have 3 candidates, however, only 1 satisfies the BER criterion of being <0.2, i.e., add12u\_0AF. 
Similarly for a power budget of <140 $\mu W$ we have 6 candidates fulfilling the criterion, however, only 2 satisfy the BER criterion of <0.2, namely, add12u\_0AF and add12u\_0ZP. 
Therefore, the 3D space presented in Figure~\ref{fig:dse_bask} presents us with an opportunity to make required trade-offs based on the designer's requirements to generate low-power Viterbi decoders. 

$\bullet$ \textbf{BPSK:} From Figure~\ref{fig:dse_bpsk}, we can analyze the design space for BPSK. Similar to BASK, if the required BER is <0.2, then we have 6 candidates satisfying the criterion. In parallel, similar observations can be made for a specific area and power budgets coupled with a BER threshold to shortlist adder candidates in order to have optimal designs in place for low-power Viterbi decoders.

$\bullet$ \textbf{QPSK:} For QPSK, Figure~\ref{fig:dse_qpsk} gives us a comprehensive understanding of the type of adders suitable for generating low-power Viterbi decoders. 
We can see that for a power budget < 130 $\mu W$, we have 4 adders satisfying the criterion, namely, add12u\_103, add12u\_0AF, add12u\_28B, add12u\_0AZ, however, if the required BER threshold <0.2 among these, only add12u\_0AF satisfies both criteria. Therefore, similar observations can be made for various combinations involving BER, area, and power based on the designer's/application's needs.

Though we have the above results as a conclusion per modulation scheme, it is important to note that this is valid only for the communication model presented in Figure~\ref{fig:overview} and its properties mentioned in Table~\ref{tab:comm_sys} and~\ref{tab:prop}. 
With a change in the communication model and its properties, \papername{} will be useful to determine the optimal design points.

\subsection{Green NLP}
\label{sec:eval:nlp}
Towards measuring the efficacy of approximations in the Viterbi decoder on a broader spectrum, we also evaluate Parts-of-Speech (POS) Tagging using Approximate Viterbi decoding to explore low-power and green NLP.
\subsubsection{Accuracy Analysis}
For accuracy analysis, 
we implement a POS Tagging system in an NLP setting with approximate ACSUs (approximate adders) using Python. 
To test our approach, we take 3 sentences containing 2, 3, and 6 words, respectively.

We evaluate 15 approximate adders from the EvoApprox Library~\cite{evoapprox} to build Approximate Viterbi decoders to tag the parts of speech in these sentences and report the accuracy level. 
Upon analysis we find that 7 adders report 100\% accuracy, namely, add16u\_1A5, add16u\_0GN, add16u\_0TA, add16u\_15Q, add16u\_162, add16u\_0NT, and add16u\_110. add16u\_0NL has an accuracy of 88.89\%. 
The rest of the adders have an accuracy below 60\%.

The results above show that for an application like POS Tagging, using approximate adders instead of accurate ones can provide an accuracy level of 100\% in many cases, thereby supporting the use of approximations.

\begin{figure}
\begin{tikzpicture}

\definecolor{darkgray176}{RGB}{176,176,176}
\definecolor{gray}{RGB}{128,128,128}
\definecolor{lightgray204}{RGB}{204,204,204}

\begin{axis}[
legend cell align={left},
legend style={
    fill opacity=0.8, 
    draw opacity=1, 
    text opacity=1, 
    at={(1.2,1.15)},
    draw=lightgray204
},
tick align=outside,
tick pos=left,
x grid style={darkgray176},
xmin=0, xmax=478.4409,
xtick style={color=black},
y grid style={darkgray176},
ymin=-0.2, ymax=17,
width=0.35\textwidth,
height=7cm,
ytick style={color=black},
ytick={0.4,1.4,2.4,3.4,4.4,5.4,6.4,7.4,8.4,9.4,10.4,11.4,12.4,13.4,14.4,15.4,16.4},
yticklabels={
  Ripple Carry Adder,
  CLA,
  add16u\_07T,
  add16u\_1A5,
  add16u\_0GN,
  add16u\_126,
  add16u\_0TA,
  add16u\_15Q,
  add16u\_162,
  add16u\_0U8,
  add16u\_0UV,
  add16u\_0VA,
  add16u\_067,
  add16u\_0NT,
  add16u\_0NL,
  add16u\_08V,
  add16u\_110
}
]
\draw[draw=none,fill=black] (axis cs:0,-0.2) rectangle (axis cs:369.47,0.2);
\addlegendimage{ybar,ybar legend,draw=none,fill=black}
\addlegendentry{Area ($\mu m^2$)}

\draw[draw=none,fill=black] (axis cs:0,0.8) rectangle (axis cs:455.658,1.2);
\draw[draw=none,fill=black] (axis cs:0,1.8) rectangle (axis cs:231.42,2.2);
\draw[draw=none,fill=black] (axis cs:0,2.8) rectangle (axis cs:366.814,3.2);
\draw[draw=none,fill=black] (axis cs:0,3.8) rectangle (axis cs:329.31,4.2);
\draw[draw=none,fill=black] (axis cs:0,4.8) rectangle (axis cs:316.008,5.2);
\draw[draw=none,fill=black] (axis cs:0,5.8) rectangle (axis cs:365.218,6.2);
\draw[draw=none,fill=black] (axis cs:0,6.8) rectangle (axis cs:361.494,7.2);
\draw[draw=none,fill=black] (axis cs:0,7.8) rectangle (axis cs:355.11,8.2);
\draw[draw=none,fill=black] (axis cs:0,8.8) rectangle (axis cs:356.706,9.2);
\draw[draw=none,fill=black] (axis cs:0,9.8) rectangle (axis cs:341.278,10.2);
\draw[draw=none,fill=black] (axis cs:0,10.8) rectangle (axis cs:319.2,11.2);
\draw[draw=none,fill=black] (axis cs:0,11.8) rectangle (axis cs:183.806,12.2);
\draw[draw=none,fill=black] (axis cs:0,12.8) rectangle (axis cs:329.308,13.2);
\draw[draw=none,fill=black] (axis cs:0,13.8) rectangle (axis cs:339.15,14.2);
\draw[draw=none,fill=black] (axis cs:0,14.8) rectangle (axis cs:312.018,15.2);
\draw[draw=none,fill=black] (axis cs:0,15.8) rectangle (axis cs:356.706,16.2);
\draw[draw=none,fill=gray] (axis cs:0,0.2) rectangle (axis cs:184.57,0.6);
\addlegendimage{ybar,ybar legend,draw=none,fill=gray}
\addlegendentry{Power ($\mu W$)}

\draw[draw=none,fill=gray] (axis cs:0,1.2) rectangle (axis cs:222.542,1.6);
\draw[draw=none,fill=gray] (axis cs:0,2.2) rectangle (axis cs:44.195,2.6);
\draw[draw=none,fill=gray] (axis cs:0,3.2) rectangle (axis cs:179.756,3.6);
\draw[draw=none,fill=gray] (axis cs:0,4.2) rectangle (axis cs:134.924,4.6);
\draw[draw=none,fill=gray] (axis cs:0,5.2) rectangle (axis cs:116.98,5.6);
\draw[draw=none,fill=gray] (axis cs:0,6.2) rectangle (axis cs:170.477,6.6);
\draw[draw=none,fill=gray] (axis cs:0,7.2) rectangle (axis cs:165.3047,7.6);
\draw[draw=none,fill=gray] (axis cs:0,8.2) rectangle (axis cs:159.27,8.6);
\draw[draw=none,fill=gray] (axis cs:0,9.2) rectangle (axis cs:155.88,9.6);
\draw[draw=none,fill=gray] (axis cs:0,10.2) rectangle (axis cs:140.341,10.6);
\draw[draw=none,fill=gray] (axis cs:0,11.2) rectangle (axis cs:133.2833,11.6);
\draw[draw=none,fill=gray] (axis cs:0,12.2) rectangle (axis cs:44.434,12.6);
\draw[draw=none,fill=gray] (axis cs:0,13.2) rectangle (axis cs:134.7229,13.6);
\draw[draw=none,fill=gray] (axis cs:0,14.2) rectangle (axis cs:146.3073,14.6);
\draw[draw=none,fill=gray] (axis cs:0,15.2) rectangle (axis cs:118.99,15.6);
\draw[draw=none,fill=gray] (axis cs:0,16.2) rectangle (axis cs:164.865,16.6);
\end{axis}

\end{tikzpicture}
    \caption{Area and Power Statistics for Parts-of-Speech Tagger (NLP) using various adders in the Viterbi Decoder}
    \label{fig:ap_nlp}
    \vspace{-3ex}
\end{figure}
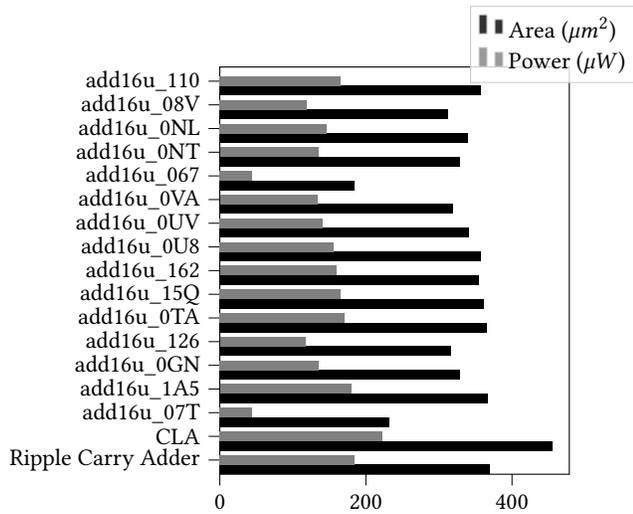

\subsubsection{Hardware Evaluation}
Similar to Section~\ref{sec:eval:comm:hw}, hardware evaluation is done using 15 16-bit approximate adders (shown in Figure~\ref{fig:ap_nlp}) from the EvoApprox Library.  
Upon analysis, we find that add16u\_07T consumes the lowest power, i.e., 44.195 $\mu W$. 
However, add16u\_07T has an accuracy of 16.663\% which is unsuitable for practical cases.
Therefore, to study accuracy v/s hardware trade-off, we explore the design space of a POS tagger with approximate ACSUs to obtain optimal decoder designs in the following subsection.

\begin{figure}
    \includegraphics[width=0.4\textwidth]{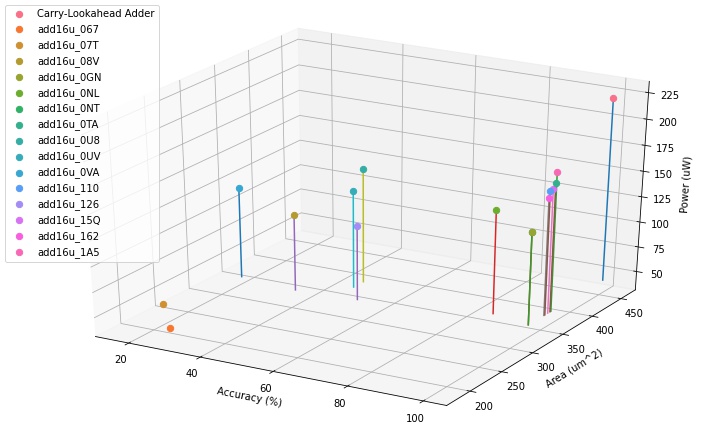}
    \caption{Design Space Exploration of Viterbi decoders for Green NLP with various adders}
    \label{fig:dse_nlp}
\end{figure}

\subsubsection{Design Space Exploration (DSE)}
\label{sec:dse_nlp}
A 3D grid having the accuracy, area, and power for each design point is shown in Figure~\ref{fig:dse_nlp}.
The goal is to find optimal design points in order to generate low-power Viterbi decoders for different use cases. 
From Figure~\ref{fig:dse_nlp}, we can conclude that for a power budget of < 120 $\mu W$, we have 4 candidates satisfying the criterion, however, none of those have an accuracy level > 60\%. 
Similar observations and trade-offs can be made based on the designer's/application's requirements for a particular accuracy level coupled with a suitable power and area budget.

\section{Conclusion and Future Work}
We presented \papername{}, a DSE approach for evaluating pareto-optimal low-power Viterbi decoders satisfying accuracy and area constraints. 
\papername{} exploits approximate adders for the addition operations inside the compute-intensive ACSU of the Viterbi decoder. 
We demonstrated the utility of \papername{} on
two exemplar applications involving Viterbi decoding: an end-to-end digital communication system and a POS tagger in NLP. 
Our experimental results demonstrate that \papername{} discovers pareto-optimal design points that meet quality constraints while reducing both on-chip area and power.

Each application domain exhibits a different sensitivity to approximations, requiring a thorough DSE using a framework such as \papername{}.
For instance, our \papername{}-generated design spaces reveal that digital communication systems are less tolerant to the effects of approximation compared to a POS  tagger in NLP. 
This variation is dependent on the communication model and the system properties involved. 
The exploration of design points is application-specific, and a generic approach may not always be optimal, as evidenced by a marginal accuracy loss of 0.142\% in Bit Error Rate (BER) for digital communication and the high accuracy achieved for POS tagger.
Future work will investigate integrating \papername{} with other complementary methods such as those described in~\cite{vit1,vit2, nitin} to further improve the efficiency of Viterbi decoding in end-to-end systems.










\bibliographystyle{unsrt}
\bibliography{main}


\end{document}